\documentstyle[aps,epsfig]{revtex}
\begin{document}
\title{A Wormhole at the core of an infinite  cosmic string}
\author{ Rodrigo O. Aros\footnote{e-mail:raros@cec.uchile.cl},
\qquad Nelson Zamorano\footnote{e-mail:nzamora@cec.uchile.cl}\\
Departamento de F\'\i sica, Universidad de Chile\\
Casilla 487-3, Santiago, Chile}

\maketitle

\abstract{
We study a solution of Einstein's equations that describes a straight cosmic 
string with a variable angular deficit, starting with a $2\,\pi$ deficit at 
the core. We show that the coordinate singularity associated to this defect
can be interpreted as a traversible wormhole lodging at the the core of the 
string. A negative energy density gradually decreases the angular deficit as 
the distance from the core increases, ending, at radial infinity, in a 
Minkowski spacetime. The negative energy density can be confined to a small 
transversal section of the string by gluing to it an exterior  Gott's like 
solution, that freezes the angular deficit existing at the matching border. The
equation  of state of the string is such that any massive particle may stay at
rest anywhere in this spacetime. In this sense this is 2+1 spacetime solution.
\par

A generalization, that includes the existence of two interacting parallel 
wormholes is displayed. These wormholes are not traversible.\par

Finally, we point out that a similar result, flat at infinity and with a $2\,
\pi$ defect (or excess) at the core, has been recently published by Dyer and 
Marleau. Even though theirs is a local string fully coupled to gravity, our toy
model captures important aspects of this solution.}

\section{Introduction and Summary}

Cosmic strings are static cylindrically symmetric objects
generated by phase transition of a self interacting scalar field minimally 
coupled to a gauge field, at the beginning of the universe. They may have
left a print behind  in the form of growing density inhomogeneities (see for 
instance \cite{rees:mnrs} or  \cite{pm:apj}). So far there is no observation 
that precludes the existence of cosmic strings in the early universe. The 
effects of these primordial inhomogeneities remain inside the allowed band of
density fluctuations established by the COBE results \cite{ssw:s}.\par

There exists several authoritatives reviews that describe the different
scenarios that give rise to the formation of cosmic strings and include the 
description of its topological and physical properties \cite{k}, \cite{v}.
For this reason we will not go through a detailed introduction here.\par

The search for an  exact analytic  solutions of a gauge field coupled to 
gravity has proved to be a hard one. An exact analytic solution for a global 
string, static and with cylindrical symmetry was found by Cohen and Kaplan 
\cite{ck:plb}. It describes the outer region of the string and, as pointed by 
Vilenkin and Shellard \cite{v}, it has a singularity at a finite radius of its
core. The linear approximation to the field equations, developed by Harari and
Sikivie \cite{h1} has been used to understand the geometry behind the Cohen
and Kaplan model. However, the study of a cosmic string using a 
linearized solution of Einstein's equations rapidly drags in a non-physical 
singularity \cite{h1}, \cite{h2} around the string.\par

Concerning the singularity mentioned above, Gregory \cite{ruth2} and later
Gibbons et al.\cite{gib}, concluded that the metric associated with a global 
string must develop a singularity at a finite distance of the string. This was 
confirmed by a numerical approach to this example developed by Laguna and 
Garfinkle\cite{laguna2}. 



Laguna-Castillo and Matzner \cite{laguna1}, using a numerical approach, have 
proved that for a wide range of the vacuum expectation value of the scalar 
field  $\eta$, with $\eta <10^{18}$ Gev, the solution at radial infinity
resembles a Minkowski spacetime minus a wedge. Laguna and Garfinkle 
\cite{laguna2}, studied a supermassive cosmic string finding Kasner-like 
metric away from the core of the string. Ortiz \cite{ortiz}, extended this 
result finding a new solution with a singularity at a finite distance from the 
core of this supermassive cosmic string. The energy density of a supermassive 
cosmic string takes the value $\mu \sim \eta^2 \approx 10^{-2}$, 
where the usual value for the symmetry breaking  energy scale is $\mu \sim 
10^{-6}$. A comprehensive and critical summary of these results appears in 
Raychaudhuri \cite{ray}.\par 

Along this line, the work of Dyer and Marleau  \cite{dyer} is the more
relevant for our results. They found strings solutions with an angle surplus 
at the core and asymptotically Minkowskian (no wedge at radial infinity) using
a numerical approach. Our work reaches a similar result, working analytically 
with a simplified model of a cosmic string.\par

The simplified approach we just mentioned refers to models where the radial 
and angular components of the stress energy tensor of the string has not been 
considered. Simultaneously, a fixed width to the string is assigned and
usually is matched to  an empty spacetime \cite{g:apj}, \cite{linet}, 
\cite{his}. This is a generalization of the local cosmic string first 
introduced by Vilenkin \cite{v}.\par

From the previous account we conclude that the search for exact solutions of 
local scalar theories coupled to gravity seems to be unsurmountable 
analytically. For this reason in this work we have adopted a simplified stress
energy tensor of the kind mentioned in the last paragraph: keeping only the 
$T_{z\,z}$ and $T_{t\,t}$ components as different from zero. This model has 
been studied previously (see for instance \cite{g:apj} \cite{his}) and has 
succeeded in raising interesting questions about the physics taking place 
around a cosmic string. However, as Raychaudhuri has showed \cite{ray}, this 
approach is not a cosmic string in the sense that the boundary conditions set 
the gauge field and the complex scalar field to zero. However, as becomes 
clear from the previous references, the existence of a model, although simple,
susceptible to an analytic study may be of great help to understand the nature 
of more complicated and realistic objects. Is in this context that the results 
exhibited by Dyer and Maleau interest to us, since they have similar boundary 
conditions at the core of the string and reach the same geometry at radial 
infinity.\par

Another recent work in the same spirit that ours belongs to Cl\'ement 
\cite{cl1}, who obtained a solution for a multiwormhole associated with a 
scalar field in 2+1 dimensions and extended it to a cylindrical traversable 
wormholes in four dimensions. In \cite{cl2}, he has been able to glue several
 cosmic strings in a flat background to form a wormhole in the spirit of the 
Wheeler-Misner \cite{mw}.\par 

In our case, we will explore the physical consequences of the existence of a 
wormhole at the core of a simplified model for a cosmic string, already 
described. Our solution arises when we allow a different boundary condition at
 the core of the string. The curvature in our solution decreases slowly from
the core of the string to the radial infinity, this a difference with 
Cl\'ement's model.\par

We display an exact solution of Einstein's equations with cylindrical symmetry.
It is asymptotically minkowskian in the radial direction and the stress energy 
tensor is such that $T_{zz}=T_{tt}$ and $T^r_r = T^{\theta}_{\theta}=0$. It has
an angular deficit at the core, that slowly vanishes as we move toward 
infinity. The radial patch of coordinates does not have a curvature singularity
in $1\le r<\infty$. At $r = 1$, it contains a logarithmic coordinate 
singularity, that we interpret as the location of a traversible wormhole 
connecting two identical universes along the string. This is the only extension
of this metric we have studied, though it is not the only one possible.\par



We made an effort to find a family of solutions that would break the 
cylindrical symmetry and, simultaneously, would show the existence of (at 
least) a couple of wormholes, keeping the asymptotic flatness of the metric. 
However, we only found a solution where the wormholes cannot be crossed 
in a finite proper time. \par

Since the source of this cosmic string has a negative energy density, and it is
advisable to keep the transversal cross section of this string  as small as 
possible  \cite{t}. We have added a solution that can be matched to the 
wormhole metric and has a positive energy density. However, for all our 
computations  we use  the original metric since we are interested in the 
physics occurring in the neighborhood of the wormhole throat.\par

It is worth to mention that since our background metric is not empty, our
solution differs from the one found by Cl\'ement \cite{cl2}.\par
 
The location of the wormhole is arbitrarily set at $r=1$. This number has no
physical meaning, we only have established an arbitrary scale for the
coordinates.\par 
 
We use the signature conventions and the definitions of the Riemann and Ricci 
tensor appearing in  Wald's book,\cite{wald}. Also $G=\hbar=c=1$. Greek
indexes run from 1...4. \par

\section{A traversible wormhole}

\subsection{The metric}

Let's consider the following metric:\par
\begin{eqnarray}
d\,s^2 = - d\,t^2 +
 \left(1 + \frac{B}{\ln\,(r/r_0)} \right) d\,r^2 + r^2\,d\,\varphi^2 +
d\,z^2\label{nelson}.
\end{eqnarray}
The only components of the Einstein's equations: 
$G_{\alpha\beta} = R_{\alpha\beta} - \frac{1}{2}\,g_{\alpha\beta}\,
R= 8\,\pi\,T_{\alpha\beta}$,
associated with this metric, setting $r_0\equiv 1$, are:\par

\begin{eqnarray}
-G_{tt} =G_{zz} = {\frac {B}{2\,r^2\,\left[ \ln (r)+B \right]^{2}\,
r^{2}}}, \qquad  T^{t}_{\,\,t}=T^{z}_{\,\,z}= -\rho,\quad \mbox{with}\quad 
\rho>0 .\label{einstein}
\end{eqnarray}

The trace of the Einstein's tensor is:
\begin{equation}
G^{\mu}_{\,\,\mu} =-R=\frac{B}{ r^{2}\,\left[\ln (r)+B \right]^{2} }. 
\end{equation}

This  solution of the Einstein's equations has a  stress energy tensor  
with pressure in the z-direction and an equation of state $p_z +\rho =0$. If 
we take $B>0$, for the reasons we state below, the energy density becomes 
negative, $\rho<0$, as seen from the equation (\ref{einstein}). Both,
the pressure and the density, decay slowly to zero at radial infinity.
As showed by Morris and Thorne \cite{t}, the existence of a wormhole 
needs a negative energy density in a region around it. As pointed out in the 
same reference \cite{t}, it is necessary to keep the negative energy density 
reduced to a small region around the wormhole. We do not ignore this here, 
however, for simplicity, all our computations adopt the metric (\ref{nelson}). 
In section IV, we will display a matching of this geometry to the Gott's 
exterior solution \cite{g:apj} to keep the negative energy density reduced to 
a small region around the wormhole.\par 

The coordinates patch valid for equation (\ref{nelson}), extends from $r =1$
to $r = \infty$, and there are no physical singularities in this region. The 
invariants associated with this metric adopt two expressions: they are 
proportional to a power of the Ricci scalar or vanish. For instance, for the 
above metric the square of the Riemann tensor is

\begin{equation}
R_{\mu\,\nu\,\sigma\,\tau}\,R^{{\mu\,\nu\,\sigma\,\tau}}
= 
\left[ B^2/{r^{4}\,\left (\ln (r)+B\right )^{4}}\right] = R^2.
\end{equation}

The Ricci scalar has singularities at $r = 0$ and  $\ln\,r =- B$, both of them
remain outside the patch of the radial coordinate $r\,\ge\,1$, if $B>0$. Thus,
under these conditions, all the  invariants are finite. For $B<0$, we have a
positive energy density and a flat spacetime at infinity, however we drag in a
naked singularity at $\ln(r)=|B|$ that hinders any physical interpretation of
the source of this metric. Note that as soon as the density changes sign the 
wormhole vanishes, as it is well known \cite{t}.\par

In the following section we will show that the coordinate singularity that
lies at $r =1$, allows us to naturally plug in another mirror spacetime there.
This throat, that connects two twin spacetimes, is what we have called here a 
wormhole.\par
 
\subsection{Geodesics}

The geodesics equations for a lightlike vector  associated with this metric 
(\ref{nelson}) are:\par

\begin{eqnarray}
- \dot{t}^2+ \left(1+ \frac{B}{\ln\,r}  \right) \,\dot{r}^2 +
-r^2 \dot{\varphi}^2 =0,&\quad&-\,\dot{t} = E,\quad \dot{\varphi}\,r^2 = L.
 \label{null}
\end{eqnarray}
Here, the prime indicates a radial derivative and the dot, a derivative with
respect to the affine parameter.  We also have assumed that the motion takes 
place on the $z=$ constant, plane. Solving these equations for $\dot{r}$, we 
have:\par
\begin{equation}
\dot{r} = \pm\,\frac{\sqrt{\left( E^2 -\frac{L^2}{r^2} \right) } }{ 
\sqrt{\left(1 + {B}/{\ln\,r} \right)} }.
 \end{equation}
The coordinate patch becomes singular at $r=1$. To study the trajectory of
test particles near the wormhole throat, we introduce a new set of coordinates
that avoids this singularity.  Defining a radial coordinate as
\begin{equation} 
\chi = \pm \int \sqrt{\left(1 + \frac{B}{\ln(r)}\right)}\,\frac{1}{r}\, dr 
\label{tra1},
\end{equation}
here each sign maps a different side of the wormhole, welding both sides at
$r=1$. The explicit dependence of $\chi $ on $r$, obtained from (\ref{tra1}), 
is:
\begin{equation}
  \chi = \pm\left[ \sqrt{\ln(r)}\sqrt{\ln(r) + B} +
B\ln\left(\sqrt{\ln(r)} + \sqrt{\ln(r)+ B} \right) - \frac{1}{2}
B\ln(B)\right],  \label{tra3}
\end{equation}
where $\chi \in (-\infty \ldots \infty)$ and  we have added a constant to set 
$\chi|_{r=1} = 0$. In these coordinates the metric takes the expression:

\begin{equation}
ds^2 = P(\chi)\left(d\chi^2 + d\phi^2\right) + dz^2 - dt^2,\quad \mbox{ with
 }\,  P(\chi)=r^2.
\end{equation}

Assuming a geodesic equation with just a radial dependence $\chi$, we obtain 
the following general solution (with $z=$constant):
\begin{equation}
l^a = \left( E,\,\pm\sqrt{\frac{D - {L^2}/{P(\chi)} }{P(\chi)}},
\frac{L}{P(\chi)}, 0 \right),\qquad \mbox{where }\quad\quad D=\left\{
 \begin{array}{lll}  E^2 &\mbox{if}&\quad l_a l^a  = 0, \\
                            E^2-1 &\mbox{if}&\quad l_a l^a =-1. 
\end{array} \right.
\label{geose}
\end{equation}

Since  there is not a closed form expression for $r= r(\chi)$, we study the 
 $r \approx 1$ region. There, the equation (\ref{tra3}) can be approximated, 
to the lowest non-trivial term, as $ \chi \approx  \pm 
2\sqrt{B}\,\sqrt{r-1}$, obtainning $ P(\chi)=r^2 \approx 1 + {\chi^2}/{2B}$.
\par

Using this approximation and the equation (\ref{geose}), the evolution of any 
kind of test particles near the wormhole throat, is given by the 4-vector:
\begin{equation}
l^a = (E,\pm \sqrt{D-L^2},L,0).
\end{equation}
This result points to the existence of particles orbiting at $\chi=0$,
if $D=L^2$. Otherwise, if $D>L^2$, the particle crosses the wormhole 
throat, going from one universe to the other. We picture this situation as the 
test particle  crossing the wormhole and appearing in a twin universe, an 
analytic continuation of the original spacetime.\par
On the opposite case, if $D<L^2$, the test particle bounces at some radius 
larger than unity ($\chi>0$), avoiding the throat.\par 

One of the geodesics that belongs to this family is pictured in Figure 2. It
corresponds to an embedding of the conical structure of the throat in a three 
dimensional space. The line threading this surface is a lightlike geodesic
obtained through a numerical computation of the orbit.\par

To prove that the wormhole is traversible, we need to compute the time a 
particle spends crossing the wormhole. In the absence of a closed form for 
$r=r(\chi)$, we use the approximation established for the neighborhood
of $r = 1$ or $\chi=0$. Let's compute the time it takes an infalling particle
to reach $r=1$ from another point close to it: 

\begin{equation}
\frac{d\,\chi}{d\,t}= \frac{1}{E}\,\sqrt{\frac{D\,P(\chi)- L^2}{
P^2(\chi)}},\quad \mbox{therefore}\quad t=-\int_{\chi}^0\,E\, \sqrt{
\frac{P^2(\chi)}{D\,P(\chi)- L^2} }\,d\chi, 
\end{equation}
the minus sign corresponds to an infalling particle. This integral can be
evaluated exactly (with the approximation $P(\chi)\approx 1+\chi^2/(2\,B) $ 
already defined). The result is finite unless $D-L^2$ vanishes. This
computation shows that this wormhole is traversible.\par 

For $D- L^2=0$, the integral is proportional to $\ln \chi$, so for an
infalling particle (massless or not) it takes an infinite time to reach the 
throat under this conditions. This result points to a possible appearance of an
instability at the throat in case that part of the background radiation from 
the surrounding  get stored, orbiting at $r=1$ or near it. We think that most 
probably the back reaction generated by this matter may close the wormhole. 
There is a hint in this direction, the spherical symmetric Lorentzian wormhole 
spacetimes has been unable to support a quantized massive scalar field 
\cite{his2}.\par

Note that a massive particle may remain at rest at any value of the $(r,\phi)$
coordinates, as it can be seen from equation (\ref{geose}). In this case we
have $E=1$, $\dot{t}=1$ and $L=0$. In this sense this string is the equivalent
of the Vilenkin cosmic string. It also reflects its 2+1 connection. \par

If we set a new radial coordinate as $R=\int_1^r d\,r\sqrt{1+ B/\ln(r)} $ we 
get:
\begin{equation}
d\,s^2= -d\,t^2+ d\,R^2+ r(R)^2\,d\,\phi^2+ d\,z^2,
\end{equation}
the particle does not feel any gravity and remains at rest. Of course we can
reach the same conclusion from $g_{t\,t}=-1$, there is not a gravitational 
potential in this metric. We also learn that as soon as we introduce a 
coordinate dependence in $g_{t\,t}$ there will appear a force and the particle 
will not be able to stand at rest.\par  

\subsection{Total Mass}

The total energy density associated with this string is:

\begin{equation}
T_{t}^t =\frac{1}{8\,\pi}\, \frac{B}{2 r^2 (B + \ln\, r)^2} =  T_{z}^z, 
\quad  T_{rr} = T_{\phi \phi} = 0.
\end{equation}

To set the  singularity inside $r =1$, we need $B>0$, as a consequence we
have a negative energy density $ \rho $. The total mass per unit length 
obtained after integrating the density from $r=1$ to $r=\infty$ is  $\mu =
 -\frac{c^2}{4\,G}$.\par

A cosmic string with a positive mass per unit of length that adopts this
density,  $\mu= \frac{c^2}{4\,G}$, represents a supermassive string 
\cite{ortiz},\,\cite{laguna2}. For this spacetime, a foliation with $t$ and $z$
constants, starts flat at the center and goes as a cylinder at infinity (see 
for instance reference \cite{v}). Using this result, we can offer an intuitive
picture of our solution as follows: we start with a cylindrical geometry at 
the throat ($r=1$), generated by a so called topological mass equal to 
$\frac{c^2}{4\,G}$, then, as we move out, there are shells of negative energy 
added. Given the absence of radial and angular component imposed in the stress 
energy tensor, the excess angle introduced by the negative mass added as we 
move out, slowly cancels the $2\pi$ defect that lies at the throat.\par

Attempting to make contact with the work of Dyer and Marleau \cite{dyer}, we
added an extra term to the temporal component of the metric: $g_{00}=-1+
[C\,\ln r]/r $. With this extra term the spacetime remains asymptotically flat,
but the energy density depends on the relative values of the constants $B$ and
$C$. In this case it is possible to adjust these constants to have a positive 
mass density in the neigborhood of $r=1$. Also the extra term added turns on 
the radial and angular component of the stress energy. Even though with this 
addition we eliminated the naked singularity mentioned in the above paragraph 
for $B<0$, we were not able to give a physical interpretation to the stress 
energy associated with this term.\par  

\section{The Klein Gordon Equation}

To study the dynamics of a scalar field in this background geometry constitutes
the first step to learn about the stability of the wormhole \cite{his2}.
Besides, some of the properties observed in this solution may reappear in other
integer spin fields. In the $(t,r,\phi)$ coordinates, the Klein Gordon 
equation  takes the following form:


where $\mu$ is defined as the mass of the scalar field. We have not included a 
wave propagating along the z-axis to keep the contact with the 2+1 spacetime. 
We concentrate ourselves in the behavior of cylindrical waves in this 
background metric. We look for solutions of the kind: 

\begin{equation}
\Psi \equiv \mbox{exp}\{ i\,[\omega\, t + m\, \phi\,]\}\, \psi_{m}(r)
\label{fdescom},
\end{equation}
with $m$ an integer. The differential equation for $\psi_{m}$ is: 
\begin{small}
\begin{equation}
\left\{ \frac{1}{r\, q(r)} \frac{\partial}{\partial r}\left( \frac{r}{q(r)}\frac{\partial}{\partial r}\right)
+  \left(\kappa^2 - \frac{1}{r^2}\,m^2 \right)\right\}\, \psi_{m}(r) = 0
\label{rumeequation}
\end{equation} 
\end{small}
where $ \kappa^2 \equiv  \omega^2 - \mu^2 $. \par
 This equation (\ref{rumeequation}), does not seem to have a global analytical
solution, however it is possible to extract some information out of it. For 
instance, for $r \rightarrow \infty$ or $q(r)\,\to\,1$, it turns to a Bessel 
differential equation,  as expected, given the cylindrical symmetry of the 
problem and its asymptotic minkowskian behavior.  Using the coordinates 
introduced in (\ref{tra1}) and (\ref{tra3}), near the throat of the wormhole, 
the Klein-Gordon equation becomes:

\begin{equation}
\left(\frac{1}{1 + \frac{\chi^2}{2\,B}}\left[\frac{\partial^2}{\partial \chi^ 2} + \frac{\partial^2}{\partial \phi^2}\right] +
\frac{\partial^2}{\partial z^2} - \frac{\partial^2}{\partial t^2} \right) \Psi = \mu^2  \Psi.  \label{KG1}
\end{equation}
Using the same decomposition of $\Psi$ as the one defined previously in 
(\ref{fdescom}), we have
\begin{equation}
\frac{d^2}{d\chi^2} \psi_{m} + \left[  \frac{\kappa^2}{2B}\,\chi^2 - m^2 + \kappa^2
\right] \psi_{m} = 0. \label{KG2}
\end{equation}

The solution of this differential equation can be expressed as a combination of
 two confluent hypergeometric functions (see for instance  \cite{murphy}):

\begin{equation}
\psi_m = e^{-{u}/{2} } \left[ A \,_1F_1(p,1/2,u) +
D\, u^{1/2}\,_1F_1(p+1/2,\frac{3}{2},u) \right]
\end{equation}
with $A$ and $D$ constants, $ u \equiv i(\kappa/\sqrt{2\,B})\, \chi^2 $ and 
\begin{equation} 
4\,p = \left( i\,\left[\frac{( \kappa^2-m^2)\,\sqrt{2\,B} }{\sqrt{\kappa^2} }
\,\right] + 1 \right).
\end{equation}
For high frequencies: $\omega^2>\mu^2$, the equation (\ref{KG2}) is similar to 
the Schr\"odinger equation for the harmonic oscillator  with an inverted 
potential $V(x)=-k\,x^2/2$, where the energy is represented by
$(\kappa^2-m^2)$. If this last expression is positive, the scalar wave crosses 
the throat and reaches the other twin universe. In the opposite case the wave 
bounces at the potential barrier, the angular momentum generates a centrifugal 
barrier that reflects the wave preventing it from crossing the throat.
Finally, for low frequency waves $\kappa^2<0$, the wave tunnels through to 
reach the other side of the throat.\par


\section{Squeezing the wormhole}

It is advisable \cite{t}  to reduce the size of the spacetime where the 
energy conditions are not obeyed. Inspired in the work of Gott \cite{g:apj}, 
we enclose this solution with an empty spacetime endowed with a conical 
singularity. We can visualize the idea as follows: keep $t$ and $z$ constants, 
and embed the remaining 2-surface in a three dimensional space, there we have 
a smooth surface with a continuously decreasing angular deficit (as displayed 
in Figure 2) and at a certain radius, namely $r=r^*$, we freeze the angular 
deficit and extend it to radial infinity with the same angular deficit. \par 

To set up this program we have found convenient to introduce a new set of 
coordinates for this metric (\ref{nelson}). The new coordinate is:\par

\begin{equation}
R = \exp\left[( \sqrt{\ln r}\, \sqrt{ \ln(r) + B})+ B\,\ln(\sqrt{\ln(r)} + 
\sqrt{\, \ln(r) + B}) \right],
\end{equation}
and the metric (\ref{nelson}), becomes:

\begin{equation}
 ds^2 = A(r) \, R^{-\lambda(r)} (dR^2 + R^2 d\phi^2) + dz^2 - dt^2 
\end{equation}
 with the functions $A(r)$ and $\lambda(r)$ defined as follows:

\begin{equation}
\lambda(r) = 2 \left[1- \left( 1 + \frac{B}{ \ln \, r}\right)^{-1/2}  
\right],\quad \mbox{and}\quad 
A(r) = \frac{r^2}{R^2 (r)} \left( 1 + \frac{B}{\ln \,r} \right)^{-1/2}, 
\end{equation} 
where $\lambda(r)$ is proportional to the difference between the so called 
topological mass that hides at $r=1$ and the amount of negative mass we added 
 till $r$.\par

At a certain radius $r \equiv r^*$, we freeze the values of both constants
$A=A(r^*)$ and $\lambda(r)=\lambda (r^*)$, and continue the metric with these 
values. The metric for $r>r^*$ becomes:
 
$$
 ds^2 = A(r^*) \, R^{-\lambda(r^*)} (dR^2 + R^2 d\phi^2) + dz^2 - dt^2, 
$$ 
which corresponds to a cosmic string metric as first introduced by Vilenkin
\cite{v}. In this way, after this matching, an outside observer will not
not be able to see the wormhole inside. As remarked by Dyer and Marleau, the 
angular surplus is not known to the rest of the universe.\par   

\section{Two Parallel Wormholes}

We have found a solution for two parallel and interconnected wormholes in
the sense of Misner-Wheeler \cite{mw}. Unfortunately, the solution found is 
not traversible as we show below.\par

We looked for solutions  with no essential singularities and that turned to a 
Minkowski spacetime at infinity . This constraint was imposed to keep the 
same boundary conditions as obtained for the previous example.\par 

Let's consider the following metric:

\begin{equation}
 ds^{2} =  - \,dt^{2}\,+\,dz^{2}\,+\,h(r,\phi)\left[dr^{2} \,+\,
r^{2}\,d\phi^{2}\right] \label{metric},
\end{equation}
where the coordinate ranges are $r\in\,]0,\infty[$\,  $t$ and $z$
$\in\, ]-\infty..\infty[$ and $\phi\, \in\, [0..2\pi\,[$. The Einstein's 
equations are:\par

\begin{eqnarray}
G^{t}_{\,t} &=& G^{z}_{\,z} = - \frac{1}{2}\,R\label{RGtt},
\end{eqnarray}     
The expression for the Ricci scalar is: 
\begin{equation}
R = \frac{-1}{h(r,\phi)}\nabla^{2}\left[\ln(h(r,\phi))\right],\quad
\mbox{where}\quad \nabla^{2} \equiv \frac{1}{r}\frac{\partial}{\partial r}
\left(r\frac{\partial}{\partial r}\right) +
\frac{1}{r^{2}}\frac{\partial^{2}}{\partial\phi^{2}}.
\end{equation}

Also, the metric (\ref{metric}) obeys the following identity, that
characterizes the 2+1 geometries:
\begin{equation}
R_{\mu\,\nu\,\alpha\,\beta}\,R^{\mu\,\nu\,\alpha\,\beta} = R^{2} \label{c2}.
\end{equation}
This result indicates that the relevant physics occurs in the $z=$ constant,
plane. To reach a Minkowski spacetime at infinity, we need: $ \lim_{r\, 
\rightarrow \,\infty} h(r,\phi) = 1$.\par

The usual approach to this problem is to concentrate all the curvature as a
singularity at the throat of the wormhole and set the rest of the (two 
dimensional) spacetime flat. This is equivalent to set $\nabla^2 h(r, \phi)=0$.
This is the approach followed, for example, by Cl\'ement \cite{cl2}.\par

Looking for a solution with two singularities that represents two parallel
wormholes, we have found an expression for $h(r,\phi)$ that fulfills (in part)
the conditions we required:
\begin{equation}
h(r,\phi) = \frac{1}{ \left(1 -  \frac{4\,a^{2}\,r^2\,\cos^2\phi + 
c^{4}}{\left(r^{2} + a^{2} \right)^{2}}\right)^{2}} \label{hr}.
\end{equation}
where $a$ and $c$ are two independent parameters. For this choice of 
$h(r,\phi)$ (\ref{hr}), the Einstein's tensor is: 
\begin{equation}
\begin{array}{ll}
G_{tt}=G_{zz}& = -8\left[ a^{2}\,r^{8} + \left(2c^{4}-4a^{4} \cos 2\phi
\right)r^{6} + 2a^2\,\left(c^{4}-2\,a^{4}\sin^2 2\phi + 3a^{4}\right)\,
r^4 \right.\\
&\\
&\left.-\left(2\,c^{4}(1-4\,\cos^2\phi) +  4\,a^4\,\cos 2\phi\right)\,a^{4}r^2
+ a^{2}(a^{4}-c^4)^2 \right]\, /\, \left(r^{2}+a^{2}\right)^{6}
\end{array}.\label{Gtt}
\end{equation}
The structure of the coordinate singularity of equation (\ref{hr}) is given
by:
\begin{equation}
 \left[r^{2} + (a^{2}-c^2)\right] \left[r^{2} + (a^{2}+c^2)\right] - 
4 r^{2}a^{2}\,\cos^2\phi =0. \label{curve} 
\end{equation}
The curve described by this equation depends on the relative values of the 
parameters $a$ and $c$. Solving equation (\ref{curve}), we obtain 
\begin{equation}
r^2=2\,a^2\,\cos 2\,\phi\,\pm\,\sqrt{4\,a^4\cos^2 2\phi -(a^4-c^4)}
\label{r2}
\end{equation}

For  $a<c$, the square root in equation (\ref{r2}) is larger than the first
 term, therefore there exists a solution for $\phi=0$ and $\phi=\pi/2$: the
curve adopts peanut shape as shown in Figure 3. If $a$ starts growing,
 we reach the equality $a=c$, here the peanut collapse to a pair of touching 
loops. It is easier to see this result from the previous equation 
(\ref{curve}): for $a=c$, there are two solutions $r=0$ for any value of $\phi$
and  $r^2= 2a^2\cos 2\phi$, valid for a limited range of $\phi$, in agreement 
with the curve displayed in Figure 4. For $a>c$, there is not solution for 
$\phi=\pi/2$ because the loops detached from each other as indicated in Figure 
5, where we have drawn this curve for $c=1$ and $a=1.01$, generating two 
strings separated by a certain distance, that mimics  two parallel wormholes.
\par
Note that the two egglike figures obtained in this case (Figure \ref{fig01}), 
turn to a pair of circles as we increase the distance that mediates between 
them. What happens in this case is the following: as we move the strings 
appart, the interaction between them becomes weaker and, when the relative
distance increases enough, both strings behave like two independent cosmic 
strings.\par 

Other condition needed for the existence of a wormhole, is the absence of 
curvature singularities in this neighborhood. This requirement is also 
satisfied by the eqs.(\ref{RGtt}) and (\ref{Gtt}), as well as the  other 
invariants (\ref{c2}).\par

It is straightforward to write the geodesic equations for this geometry. Since
they do not have an analytic solution, in Figure \ref{fig2}, we have plotted
a family of infalling lightlike geodesics which are purely radial at $r=3$. We 
have used $a= 1$ and $c =1.1$ to characterize the coordinate singularity and 
we have embedded the geodesics in a three dimensional space. The geodesics 
displayed in Figure \ref{fig2}, yields information about the geometry of this
spacetime near the coordinate singularity. \par

Given the bifocal  symmetry  of the problem, there only exist  geodesics 
independent of the coordinate  $\phi$, for those cases where $\dot{\phi}=0$ 
and $\partial h(r,\phi)/ \partial \phi=0$, as it can be shown from the
geodesic equations. These conditions are saved precisely for $\phi =
\frac{n\pi}{2}$. To prove that in this case it takes an infinite time to reach
the throat, we compute the time spent for a particle radially infalling. The
radial velocity is given by $$
\frac{d\,r}{d\,t} \propto \frac{1}{\sqrt{h(r,\phi=0)} },\qquad\mbox{ therefore
}\qquad
t\propto \int\,\frac{d\,r}{[r^2- (2a^2+\sqrt{3a^4-c^4}]},
$$
extracting the singular term from this integral, we obtain
\begin{equation}
t\propto
\,\ln\left( r- \sqrt{2a^2-\sqrt{3a^4-c^4}}\right). 
\end{equation}
The border of the throat is characterized by the vanishing of the expression
between the parenthesis, therefore it becomes clear that it takes an infinite 
time to reach the throat of this wormhole for this particular case. We assume
that the rest of the orbits behave in a similar way. \par

\section{Conclusions and final remarks}

We have studied a solution of Einstein's equations with cylindrical symmetry 
with a wormhole at its core. The negative energy density region associated
with this wormhole has been minimized and confined inside a cylinder around 
the core, properly matching it to an exterior Gott's like metric. In this way
we have hidden the wormhole away from an observer at infinity. Nevertheless, 
we used the negative energy density solution throughout this work to benefit 
from its simplicity and because our main interest has been the geometry in the
 neighborhood of the wormhole.\par

The continous  change that experiments the value of the angular deficit as we 
move with respect to the core of the string, allows the presence of geodesics 
that go around the string. Some of them bounce back to infinity, if $D<L^2$, 
as if there existed a centrifugal barrier, or cross  the wormhole through, if 
$D>L^2$. In this last case, the time it takes to cross it, is finite, either 
for a lightlike or a timelike particle. There exist closed orbits at $r=1$ if 
$D=L^2$. For these particular conditions, the time it takes to reach the 
throat is infinite, therefore the particle can be considered as effectively 
trapped in the neighborhood of the wormhole. This is a source of a classical 
type of inestability, as opposed to the quantum one mentioned in the
text, in the following sense: the concentration of many particles bounded to 
reach the throat will change the sign of the energy density evaporating the 
wormhole from this region.\par

The existence of a solution to the Klein Gordon equation in the neigborhood
of the throat, opens up the possibility to study  the stability of this 
wormhole in a more quatitative way.\par

After building consistently two parallel wormholes in the same spacetime, we
discover that they are not traversible. It takes an infinite time to get close 
to the throat. It seems that the negative energy stored in the region between 
the strings streches to infinity the path to the throat. The interaction 
between these two strings is seen through the shape of the cross sections of 
the strings we have drawn in the Figures \ref{fig00}, \ref{fig02} and 
\ref{fig01}. When the wormholes are separated the interaction decreases and 
each of them get the appearence of a single isolated one.\par


A solution of the kind we have displayed here has been mentioned in the work 
of Dyer and Marleau \cite{dyer}. They have solved numerically the full set of 
equations of the global string coupled to gravity. By relaxing the boundary 
conditions at the core, they have seen strings solutions that have an angular 
surplus (deficit) at the axis and are asymptotically Minkowski at infinite. 
We claim that our solution constitues a simple  analytic partner of theirs.\par

The work of Cl\'ement \cite{cl2} is similar to ours, but the curvature appears
as a singularity at the union between the flat space and the mouth of the 
wormhole. In our case, the curvature is  smooth and everywhere different from 
zero.\par


\section{Acknowledgments}

The work of N.Z. has been partially supported by Fondecyt through the project 
\# 1950271 and DTI, project \# E-3646-9422. R.A. has been supported by a fellowship from Conicyt and Fondecyt,
project \#2970014. We would like to thank the help received from R. Soto and
A. Meza at different stages of this work.\par

\newpage

{\bf FIGURE CAPTIONS}\par

{\bf Figure 1}:\par
Here appears the matching of the two spacetime connected through $r=1$.
Each of the surfaces is an embedding on a three dimensional space of the
coordinates $r $ and $\phi$ of the wormhole.\par

{\bf Figure 2:}\par
The trajectory of a null geodesic is depicted here. The angular momentum is
smaller then the energy $ L< E $, therefore this geodesic crosses the throat of
the wormhole as shown by our calculations.\par

{\bf Figure 3:}\par

The shape of the singularity for the case $a=1$ and $c=1.1$.
The abscissa corresponds to $r\,\cos\phi$ and the ordenate to $r\,\sin\phi$.
\par

{\bf Figure 4:}\par

The loops that appear in the Figure corresponds to $a=c$. In particular, for
this case we have used $a=1$. The point where the loops converge $r=0$, is 
singular.
\par

{\bf Figure 5:}\par

The more interesting case correponds to $a>c$. For this relative values of $a$
and $c$, we obtain two disconnected loops that represent two parallel 
wormholes. The Figure corresponds to $a=1.01$ and $c=1$.\par

{\bf Figure 6}:\par

This Figure represents a family of radially infalling lightlike geodesics. We 
notice that, in their fall they mimic the wormhole throat shape.
Also we observe that the ones located at positions where the wormhole throat
is faced symetrycally remain radial along the throat.


\newpage

\begin{figure}[htb]
\centerline{\psfig{file=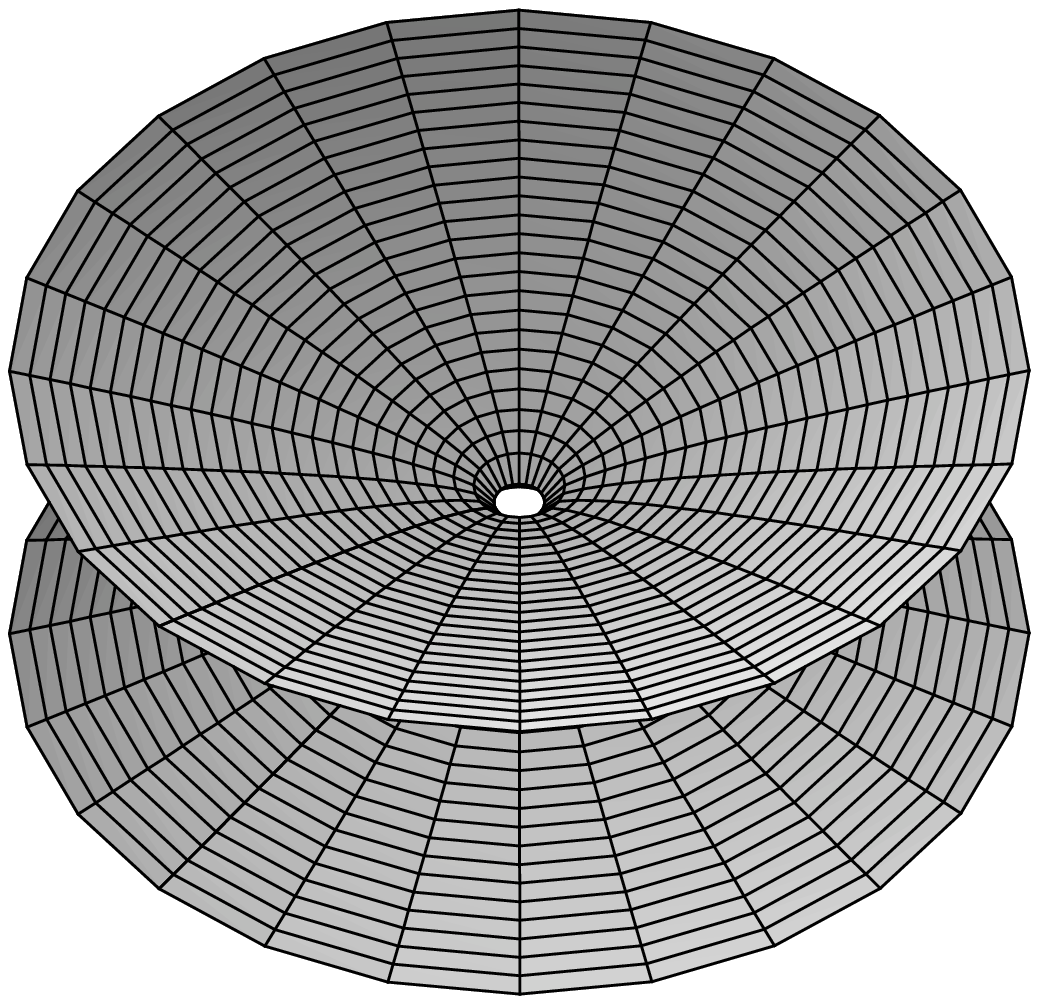,width=0.6\hsize,angle=0}}
\caption{}
\label{wormhole}
\end{figure}

\begin{figure}[htb]
\centerline{\psfig{file=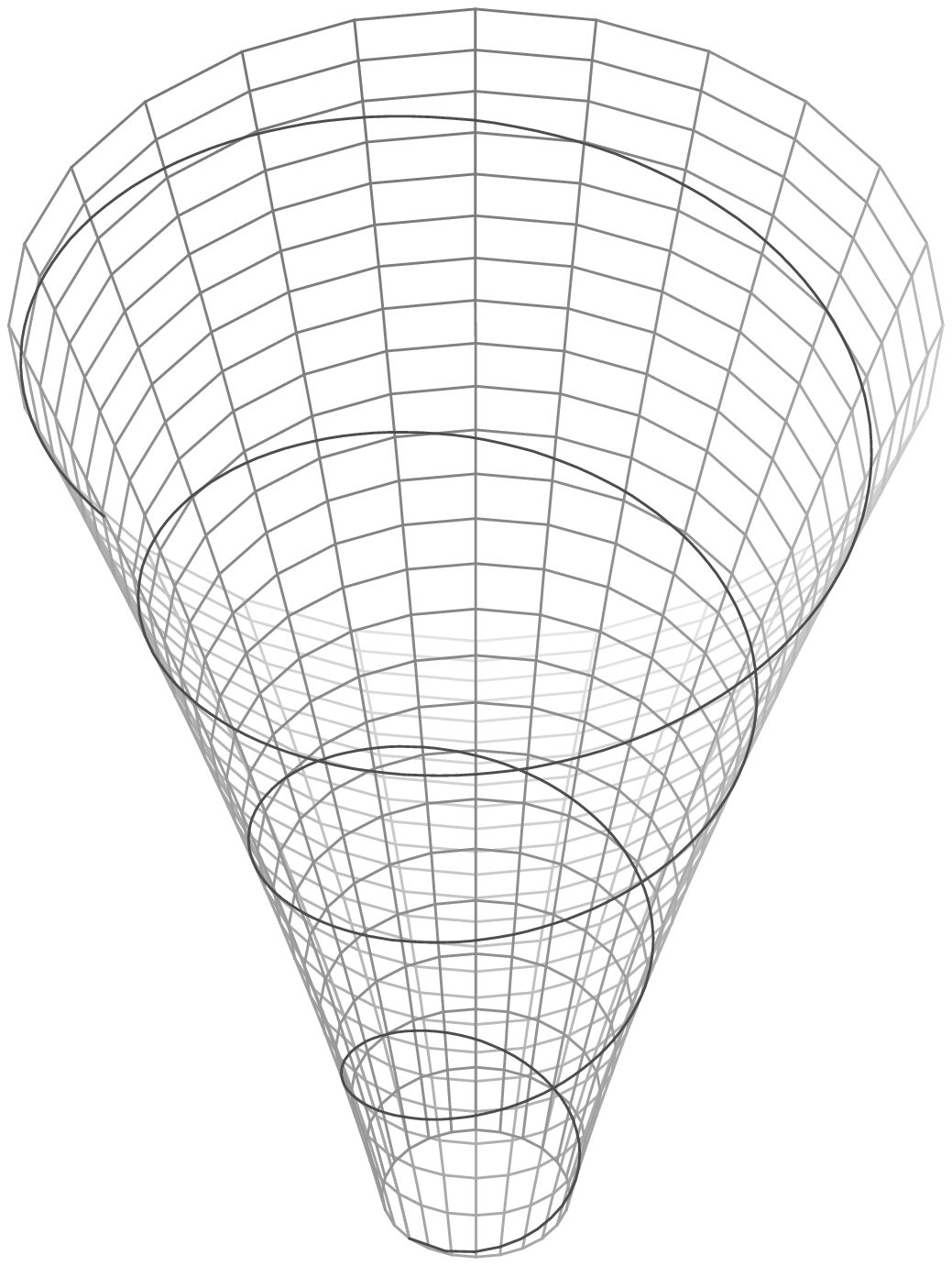,width=0.6\hsize,angle=0}}
\caption{ }
\label{wormholegeo}
\end{figure}

\begin{figure}[htb]
\centerline{\psfig{file=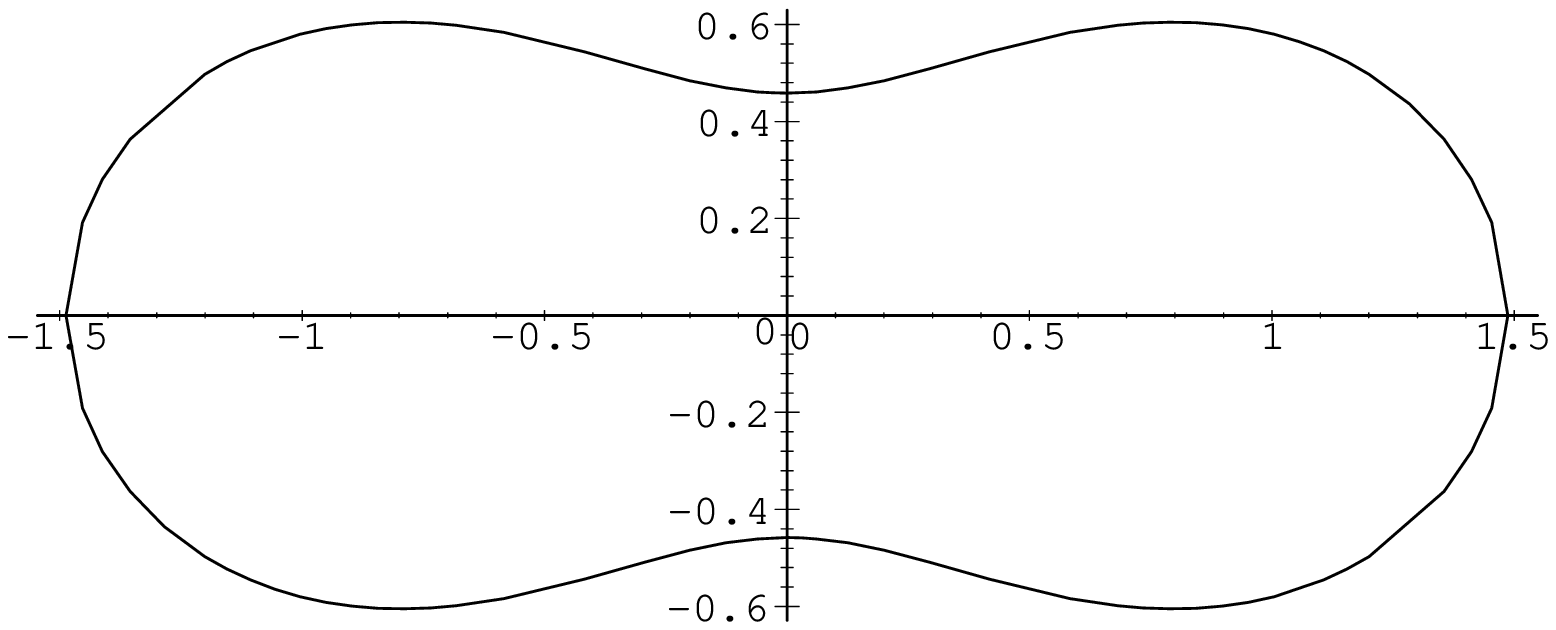,width=0.6\hsize,angle=0}}
\caption{}
\label{fig00}
\end{figure}   

\begin{figure}[htb]
\centerline{\psfig{file=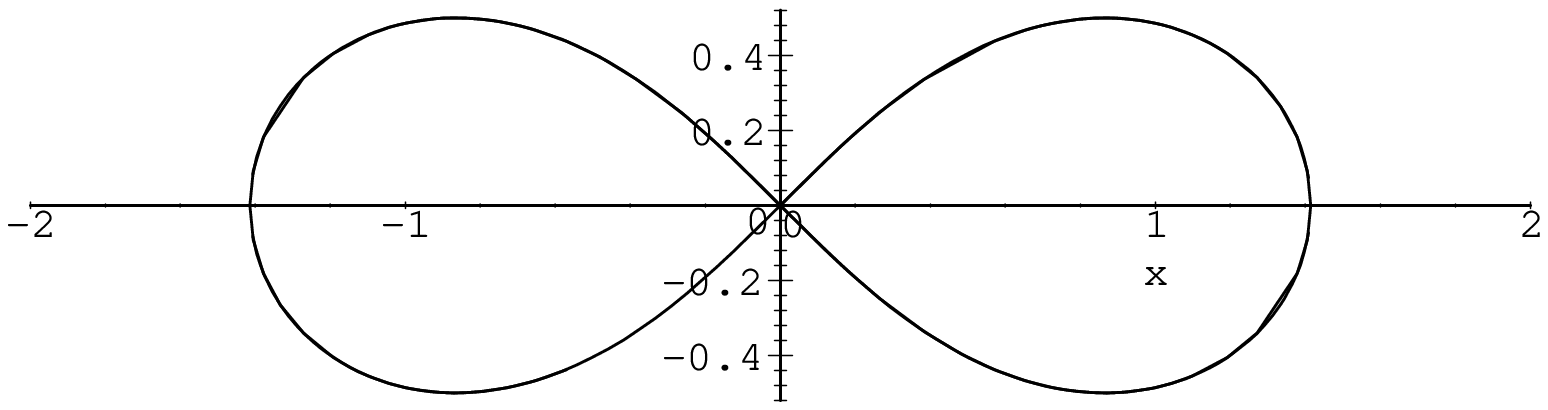,width=0.6\hsize,angle=0}}
\caption{
}
\label{fig02}
\end{figure}
   
\begin{figure}[htb]
\centerline{\psfig{file=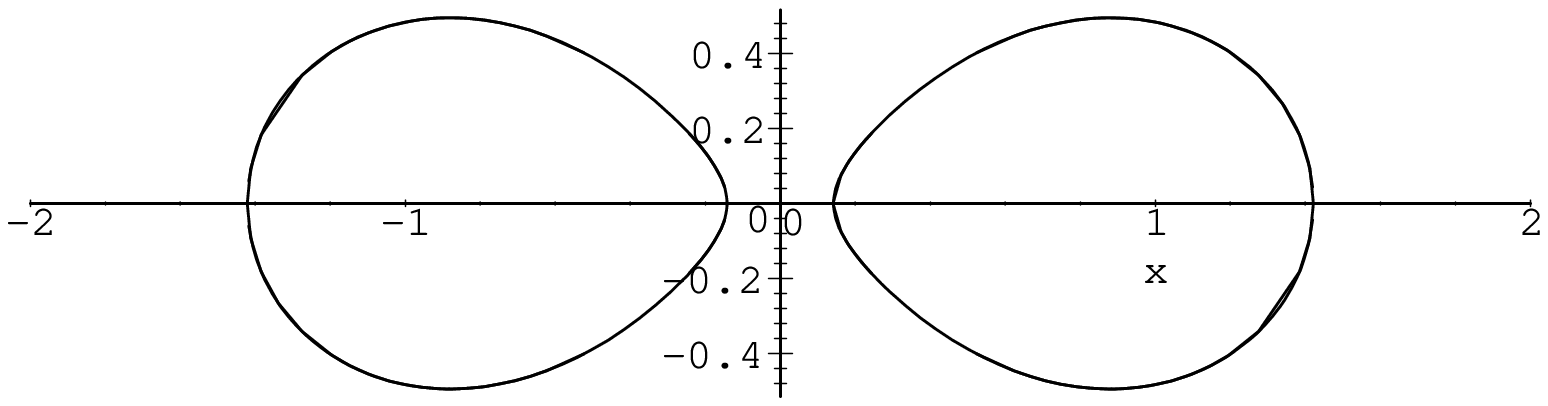,width=0.6\hsize,angle=0}}
\caption{}
\label{fig01}
\end{figure}

\begin{figure}[htb]
\centerline{\psfig{file=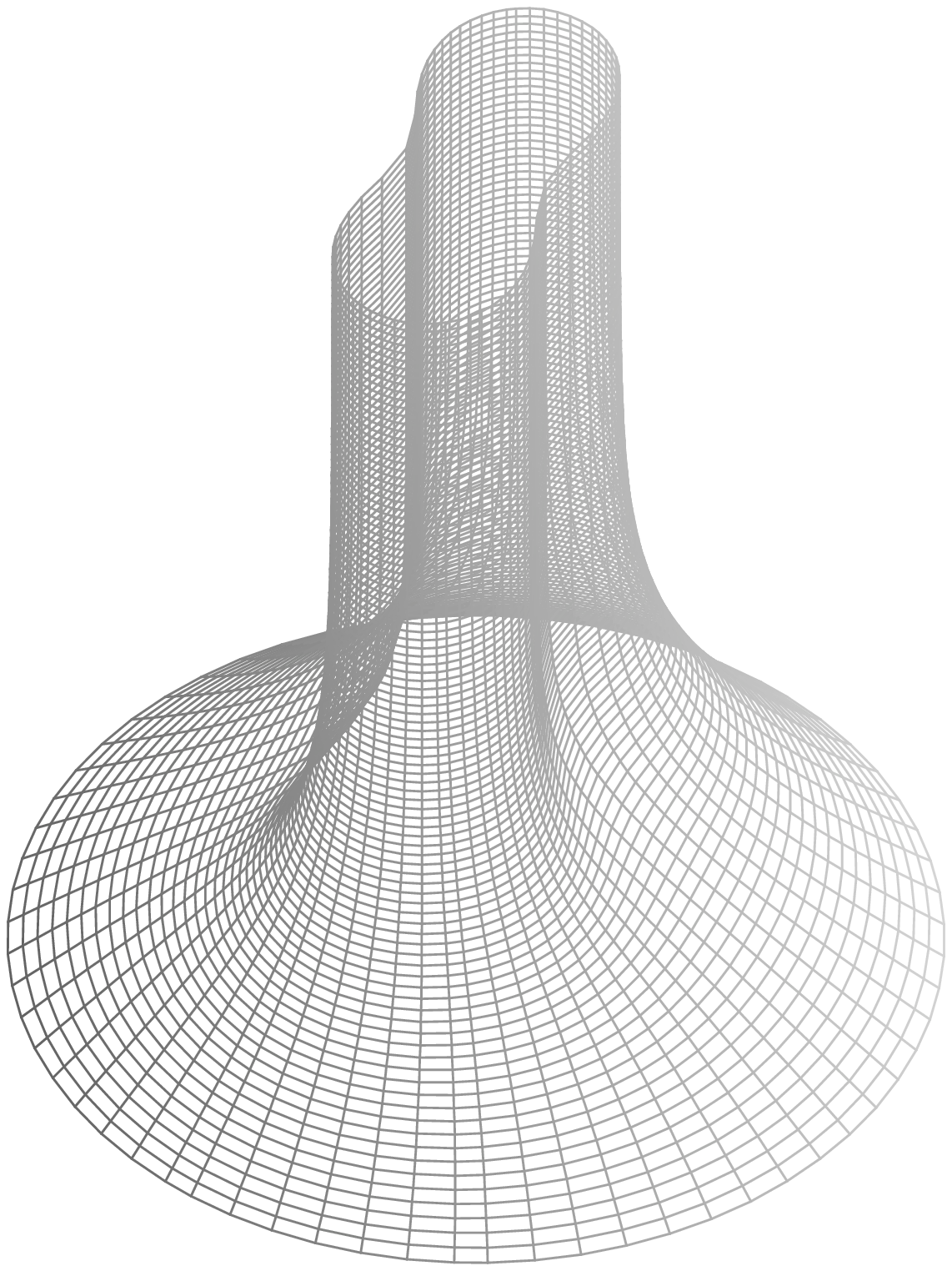,width=0.6\hsize,angle=0}}
\caption{ }
\label{fig2}
\end{figure}


\begin{thebibliography}{99}
\bibitem{units} The units of $\lambda$ must be the same used to define the
coordinate $r$. There is a scale invariance here, as noticed before.
\bibitem{inv} Due to the symmetry of the metric (\ref{metric}), its
 invariants  adopt two expressions: they are proportional to a
power of the Ricci scalar or vanish.
\end{thebibliography}
\end{document}